\documentclass[10pt,fleqn,twocolumn]{article}

\usepackage{graphicx}
\usepackage{amsmath,amssymb}
\usepackage{cite}
\usepackage[blocks,auth-sc,affil-sl]{authblk}
\usepackage[top=0.75in, bottom=0.75in, left=0.75in, right=0.75in, dvips]{geometry}

\begin{document}

\title{Exploring the Spectrum of Heavy Quarkonium Hybrids with QCD Sum Rules.}

\author{R. T. Kleiv\thanks{Speaker. E-mail: \texttt{robin.kleiv@ufv.ca}.}}
\author{B. Bulthuis} 
\author{D. Harnett}
\author{T. Richards}
\affil{Department of Physics, University of the Fraser Valley, Abbotsford, BC, V2S~7M8, Canada.}
\author{Wei Chen}
\author{J. Ho}
\author{T. G. Steele}
\affil{Department of Physics and Engineering Physics, University of Saskatchewan, Saskatoon, SK, S7N~5E2, Canada.}
\author{Shi-Lin Zhu}
\affil{Department of Physics and State Key Laboratory of Nuclear Physics and Technology, Peking University, Beijing 100871, China.}

\maketitle

\begin{abstract} 
QCD Laplace sum rules are used to calculate heavy quarkonium (charmonium and bottomonium) hybrid masses in several distinct $J^{PC}$ channels. Previous studies of heavy quarkonium hybrids did not include the effects of dimension-six condensates, leading to unstable sum rules and unreliable mass predictions in some channels. We have updated these sum rules to include dimension-six condensates, providing new mass predictions for the spectra of heavy quarkonium hybrids. We confirm the finding of other approaches that the negative-parity $J^{PC}=\left(0,1,2\right)^{-+},\,1^{--}$ states form the lightest hybrid supermultiplet and the positive-parity $J^{PC}=\left(0,1\right)^{+-},\,\left(0,1,2\right)^{++}$ states are members of a heavier supermultiplet. Our results disfavor a pure charmonium hybrid interpretation of the $X(3872)$, in agreement with previous work.
\end{abstract}


\section{Introduction}

A collection of heavy quarkonium-like XYZ states have been observed by a number of experiments (for recent reviews, see Refs.~\cite{Brambilla:2014jmp,Agashe:2014kda,Chen:2014fza,Bodwin:2013nua,Faccini:2012pj,Brambilla:2010cs}). It is difficult to understand these states within the quark model~\cite{Godfrey:1985xj}, wherein hadrons are either baryons $\left(qqq\right)$  or mesons $\left(q\bar q\right)$. Quantum chromodynamics (QCD) suggests that a far richer spectrum of hadrons is possible, including so-called exotic hadrons such as hybrid mesons $\left(qg\bar q\right)$. These are composed of a color-octet quark-antiquark pair and an excited gluon. Hybrids and mesons can have common $J^{PC}$ quantum numbers, and, because of their additional gluonic degrees of freedom, hybrids can also have exotic quantum numbers $\left(J^{PC}=0^{--},0^{+-},1^{-+},2^{+-},\ldots\right)$ which are inaccessible to mesons. Heavy quarkonium hybrids, which we shall refer to as heavy hybrids, have been studied using the constituent gluon model~\cite{Horn:1977rq}, the flux tube model~\cite{Barnes:1995hc}, the quasi-gluon model~\cite{Guo:2008yz} and lattice QCD~\cite{Liu:2012ze,Perantonis:1990dy,Juge:1999ie,Liu:2005rc,Luo:2005zg,Liu:2011rn}. These studies broadly agree that heavy hybrids should exist in the same mass range as heavy quarkonia, suggesting that some of the XYZ states could be heavy hybrids. 

QCD sum rules (QSR) is a non-perturbative method that can be used to predict hadron parameters, including masses (see Ref.~\cite{Narison:2002pw} for a comprehensive review). Refs.~\cite{Govaerts:1984hc,Govaerts:1985fx,Govaerts:1986pp} comprise the earliest QSR studies of heavy hybrids. The QSR analyses therein included leading-order perturbative and dimension-four gluon condensate contributions for several distinct $J^{PC}$ channels. However, some of the sum rules exhibited instabilities, leading to unreliable mass predictions. Recently, the sum rules for the $1^{--}$~\cite{Qiao:2010zh} and $0^{-+}$~\cite{Harnett:2012gs} channels have been updated to include the effect of dimension-six condensates. These contributions stabilize the sum rules, permitting reliable heavy hybrid mass predictions to be made. In Ref.~\cite{Berg:2012gd}, the $1^{++}$ channel was also updated, confirming previous results \cite{Govaerts:1984hc,Govaerts:1985fx,Govaerts:1986pp} for this channel. A comprehensive QSR study of these and all remaining channels was performed in~\cite{Chen:2013zia}, thereby providing updated mass predictions for the spectra of charmonium hybrids and bottomonium hybrids. The remainder of this article will discuss the QSR analysis performed in Ref.~\cite{Chen:2013zia} and its implications for the XYZ states.


\section{Laplace Sum Rules for Heavy Hybrids}

The correlation function used in the QSR analysis of heavy hybrids is
\begin{gather}
\Pi_{\mu\nu}\left(q\right) = i \int d^4 x \, e^{i q \cdot x} \langle 0 | T \left[\right. J_\mu\left(x\right) J_\nu^\dag\left(0\right) \left.\right] | 0 \rangle \,,
\label{correlator}
\end{gather}
where $J_\mu$ is an interpolating current that couples to heavy hybrids. The currents considered in Ref.~\cite{Chen:2013zia} and the corresponding hybrid quantum numbers that they couple to are as follows:
\begin{gather}
\begin{array}{ll}
J_\mu= g \bar{Q} \frac{\lambda^a}{2} \gamma^\nu G^a_{\mu\nu}Q							& J^{PC}=1^{-+},0^{-+}, \\
J_\mu= g \bar{Q} \frac{\lambda^a}{2} \gamma^\nu\gamma_5 G^a_{\mu\nu}Q					& J^{PC}=1^{+-},0^{--}, \\
J_{\mu\nu}=g \bar{Q} \frac{\lambda^a}{2} \sigma^\alpha_\mu \gamma_5 G^a_{\alpha\nu}Q	& J^{PC}=2^{-+},\ldots,
\end{array}
\label{currents}
\end{gather}
where $Q$ denotes a heavy (charm or bottom) quark, $g=\sqrt{4\pi\alpha}$ is the strong coupling, $\lambda^a$ are the Gell-Mann matrices, $\sigma^{\mu\nu}=\frac{i}{2}\left[\gamma^\mu\,,\gamma^\nu\right]$ and $G^a_{\mu\nu}$ is the gluon field strength. When the gluon field strength is replaced with its dual, $\tilde{G}^a_{\mu\nu}=\frac{1}{2}\epsilon_{\mu\nu\alpha\beta}G^{\alpha\beta}_a$, heavy hybrid states with the opposite parity are probed. Note that the current $J_{\mu\nu}$ also couples to spin-0 and spin-1 states; however, those contributions were not considered in Ref.~\cite{Chen:2013zia}. Contributions from each $J^{PC}$ channel in~\eqref{currents} can be calculated by constructing appropriate projections of the correlation function~\eqref{correlator}. Each of these projections satisfies a dispersion relation of the form
\begin{gather}
\begin{split}
\Pi\left(Q^2\right) = \left(-Q^2\right)^n &\int\limits_{4m^2}^\infty \frac{\rho\left(s\right)}{s^n\left(s+Q^2\right)}\,ds 
\\
&+ \sum_{k=0}^{n-1} a_k\left(-Q^2\right)^k \,,
\end{split}
\label{dispersion}
\end{gather}
where $Q^2=-q^2$ is the Euclidean momentum, $m$ is the heavy quark mass, $\rho\left(s\right)$ is the hadronic spectral function and $a_k\left(-Q^2\right)$ are subtraction terms. The  QCD correlation function $\Pi\left(Q^2\right)$ is related to the hadronic spectral function via
\begin{gather}
{\rm Im}\,\Pi\left(s\right)=\frac{1}{\pi}\rho\left(s\right) \,.
\label{duality}
\end{gather}
In QCD Laplace sum rules, a Borel transform is applied to both sides of~\eqref{dispersion}, which has the effect of suppressing excited state contributions in $\rho\left(s\right)$ in addition to removing the subtraction terms. The hadronic spectral function is expressed in terms of resonance and continuum contributions,
\begin{gather}
\rho\left(s\right) = \rho_{\rm res}\left(s\right) + \theta\left(s-s_0\right) \rho_{\rm cont}\left(s\right) \,,
\label{spectral_function}
\end{gather}
where $s_0$ is the continuum threshold. In Ref.~\cite{Chen:2013zia}, a single narrow resonance model was used:
\begin{gather}
\rho_{\rm res}\left(s\right) = f^2 M^{8} \delta\left(s-M^2\right) \,,
\label{resonance}
\end{gather}
where $M$ is the mass of the lowest lying heavy hybrid resonance in the $J^{PC}$ channel under consideration and $f$ is the coupling of the resonance to the heavy hybrid current. In this way, the correlation function~\eqref{correlator} was used to predict the masses of heavy hybrids.

\begin{figure}[htb]
\includegraphics[width=0.23\textwidth]{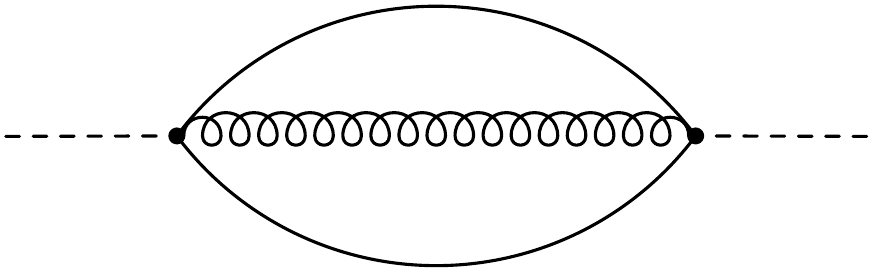}\quad\includegraphics[width=0.23\textwidth]{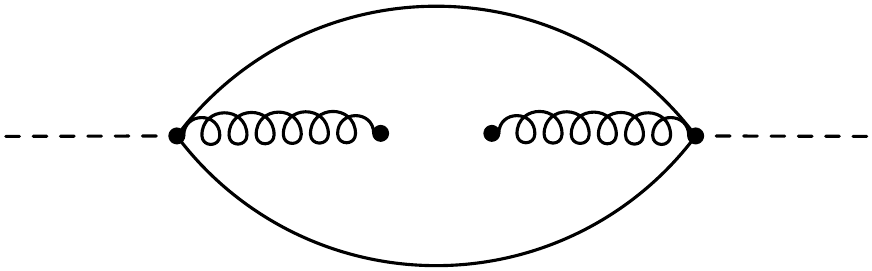}
\includegraphics[width=0.23\textwidth]{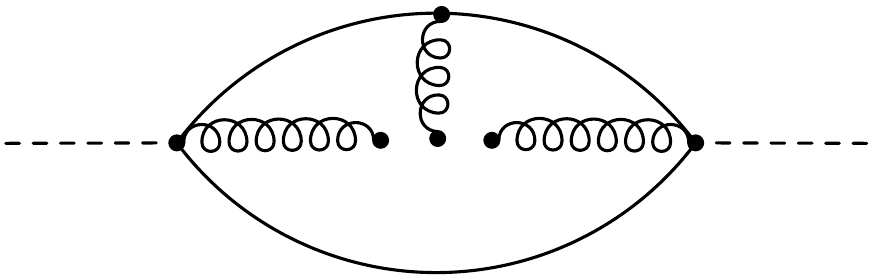}\quad\includegraphics[width=0.23\textwidth]{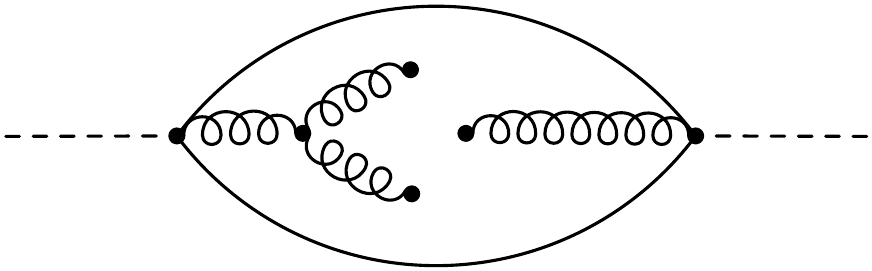}
\caption{Feynman diagrams representing contributions to the heavy hybrid correlation function. The top left and top right are the perturbative and dimension-four gluon condensate $\langle \alpha G^2 \rangle$ contributions. The bottom left and bottom right are distinct contributions of the dimension-six gluon condensate $\langle g^3 G^3 \rangle$. Dashed, curly and straight lines respectively represent the heavy hybrid current, a gluon propagator and a heavy quark propagator, whereas dots represent a condensate.}
\label{diagrams}
\end{figure}

In QSR analyses, the correlation functions are calculated using the operator product expansion. For the heavy hybrid correlation function~\eqref{correlator}, this gives
\begin{gather}
\begin{split}
\Pi_{\mu\nu}\left(q\right) = C^{\rm I}_{\mu\nu}\!\left(q\right) &+ C^{\rm GG}_{\mu\nu}\!\left(q\right)\! \langle \alpha G^2 \rangle 
\\
&+ C^{\rm GGG}_{\mu\nu}\!\left(q\right)\! \langle g^3 G^3 \rangle \ldots \,,
\end{split}
\label{ope}
\end{gather}
where $C_{\mu\nu}\left(q\right)$ are Wilson coefficients, 
\begin{gather}
\langle \alpha G^2 \rangle = \langle \alpha G^{a}_{\mu\nu}G_{a}^{\mu\nu} \rangle 
\label{dim4gluoncond}
\end{gather}
is the dimension-four gluon condensate,
\begin{gather}
\langle g^3 G^3 \rangle= g^3 f_{abc} \langle  G^a_{\mu\nu} G^b_{\nu\rho} G^c_{\rho\mu} \rangle
\label{dim6gluoncond}
\end{gather}
is the dimension-six gluon condensate and the dots represent higher dimensional terms. Feynman diagrams representing the contributions to \eqref{ope} are included in Fig.~\ref{diagrams}. There are two distinct contributions to the dimension-six gluon condensate. The first is represented by the bottom left diagram in Fig.~\ref{diagrams}. The second is represented by the bottom right diagram in Fig.~\ref{diagrams} and arises in the fixed-point gauge expansion of the dimension-four gluon condensate. Note that there is also a contribution from the dimension-six quark condensate. However, in Ref.~\cite{Chen:2013zia}, this was found to have little effect on the QSR analysis, so, for brevity, it will not be discussed here. Note that only $C^{\rm I}_{\mu\nu}\!\left(q\right)$ and $C^{\rm GG}_{\mu\nu}\!\left(q\right)$ were calculated in Refs.~\cite{Govaerts:1984hc,Govaerts:1985fx,Govaerts:1986pp}; in Ref.~\cite{Chen:2013zia}, those results were confirmed. In~\cite{Chen:2013zia}, $C^{\rm GGG}_{\mu\nu}\left(q\right)$ was calculated for all channels listed in 
\eqref{currents}. It was found that these contributions stabilize sum rules that were unstable in the original analyses~\cite{Govaerts:1984hc,Govaerts:1985fx,Govaerts:1986pp}. In Ref.~\cite{Chen:2013zia}, the Wilson coefficients~\eqref{ope} were calculated using two different approaches. First, the Mathematica package Tarcer~\cite{Mertig:1998vk} was used to reduce the number of loop integrals to be calculated. These were then calculated using results given in Refs.~\cite{Boos:1990rg,Davydychev:1990cq,Broadhurst:1993mw}, resulting in very compact expressions for the Wilson coefficients in terms of generalized hypergeometric functions. Second, a heavy quark propagator containing gluonic contributions was used to obtain integral representations of the Wilson coefficients. Complete agreement was found between these two approaches. The explicit expressions for the Wilson coefficients in terms of generalized hypergeometric functions and integral representations are tabulated in Ref.~\cite{Chen:2013zia}.


\section{Heavy Hybrid Mass Predictions}

Applying the Borel transform $\hat B$ to the dispersion relation~\eqref{dispersion} and using the single narrow resonance model~\eqref{resonance} leads to the following family of QCD Laplace sum rules:
\begin{gather}
\begin{split}
\mathcal{R}_k\left(s_0\,,\tau\right) &= f^2 M^{8+2k} e^{-\tau M^2} 
\\
&= \, \frac{\hat{B}}{\tau}\left[\left(-Q^2\right)^k \Pi\left(Q^2\right)\right]
\\
&\quad-\frac{1}{\pi} \int\limits_{s_0}^{\infty} s^k \, {\rm Im}\, \Pi\left(s\right)  e^{-\tau s} ds 
\,,
\end{split}
\label{sum_rules}
\end{gather}
where $\tau$ is the Borel parameter. The heavy hybrid mass $M$ can be determined from \eqref{sum_rules} via
\begin{gather}
M = \sqrt{\frac{R_1\left(s_0\,,\tau\right)}{R_0\left(s_0\,,\tau\right)}} \,.
\label{hadron_mass}
\end{gather}
Care must be taken when using~\eqref{hadron_mass} to determine the heavy hybrid mass $M$. A stability region where the mass is insensitive to variations in $\tau$ must first be located. Next, a range of suitable values for $\tau$ must determined: $\tau$ must be small enough to ensure convergence of the operator product expansion~\eqref{ope} and large enough to suppress excited state contributions to the sum rule~\eqref{sum_rules}. Note that, in some channels, the sum rules derived in original QSR studies of heavy hybrids in Refs.~\cite{Govaerts:1984hc,Govaerts:1985fx,Govaerts:1986pp} did not exhibit a stability region. However, in Refs.~\cite{Qiao:2010zh,Harnett:2012gs,Berg:2012gd,Chen:2013zia} it was found that when the dimension-six gluon condensate is included a stability region can be found and reliable heavy hybrid mass predictions can be made. 

One-loop $\overline{\rm MS}$ expressions for the running coupling and heavy quark mass were used in the charmonium and bottomonium hybrid QSR analyses in Ref.~\cite{Chen:2013zia}:
\begin{gather}
\begin{split}
&\alpha\left(\mu\right) = \frac{\alpha\left(M_\tau\right)}{1+\frac{25\alpha\left(M_\tau\right)}{12\pi}\log{\left(\frac{\mu^2}{M_\tau^2}\right)}} \,,
\\
&m_c\left(\mu\right) = \overline{m}_c \left(\frac{\alpha\left(\mu\right)}{\alpha\left(\overline{m}_c\right)}\right)^{12/25} \,;
\\
&\alpha\left(\mu\right) = \frac{\alpha\left(M_Z\right)}{1+\frac{23\alpha\left(M_Z\right)}{12\pi}\log{\left(\frac{\mu^2}{M_Z^2}\right)}} \,,
\\
&m_b\left(\mu\right) = \overline{m}_b \left(\frac{\alpha\left(\mu\right)}{\alpha\left(\overline{m}_b\right)}\right)^{12/23} \,.
\end{split}
\end{gather}
The numerical values of all parameters used in both the charmonium and bottomonium hybrid QSR analyses are listed in Table~\ref{parameters}. The heavy hybrid mass predictions made in Ref.~\cite{Chen:2013zia} are listed in Table~\ref{masses}. Uncertainties were determined by varying the QCD parameters in Table~\ref{parameters} around their central values.

\begin{table}[htb]
\centering
  \begin{tabular}{lll}
  Parameter					& Value											& Source(s) 													\\
  \hline
  						&												&														\\
  $\alpha\left(M_\tau\right)$ 		& $0.33$ 											& \cite{Agashe:2014kda}											\\
  $\overline{m}_c$				& $\left(1.28 \pm 0.02\right) {\rm GeV}$ 						& \cite{Agashe:2014kda,Chetyrkin:2009fv,Kuhn:2007vp,Narison:2010cg}			\\
  $\alpha\left(M_Z\right)$			& $0.118$											& \cite{Agashe:2014kda}											\\
  $\overline{m}_b$				& $\left(4.17 \pm 0.02\right) {\rm GeV}$ 						& \cite{Agashe:2014kda,Chetyrkin:2009fv,Kuhn:2007vp,Narison:2011rn,Narison:2010cg}	\\
  $\langle \alpha \, G^2 \rangle$		& $\left(7.5 \pm 2.0\right)\times 10^{-2}{\rm GeV}^4$				& \cite{Narison:2011rn,Narison:2010cg}								\\
  $\langle g^3G^3\rangle$ 		& $-\left(8.2\pm 1.0\right){\rm GeV^2}\langle \alpha \, G^2\rangle$		& \cite{Narison:2011rn,Narison:2010cg}								\\
 						&												& 														\\
  \hline
  \end{tabular}
\caption{Numerical values of QCD parameters used in QSR analyses of heavy hybrids. Values are taken from the references listed in the rightmost column.}
\label{parameters}
\end{table}

\begin{table}[htb]
\centering
  \begin{tabular}{lrr}
  $J^{PC}$			& Mass of $cg\bar{c}$ $\left({\rm GeV}\right)$	& Mass of $bg\bar{b}$ $\left({\rm GeV}\right)$ 	\\
  \hline
				&							& 									\\
 $1^{--}$			& $3.36\pm0.15$					& 	$9.70\pm0.12$						\\
 $0^{-+}$			& $3.61\pm0.21$					& 	$9.68\pm0.29$						\\
 $1^{-+}$			& $3.70\pm0.21$					& 	$9.79\pm0.22$						\\
 $2^{-+}$			& $4.04\pm0.23$					& 	$9.93\pm0.21$						\\
 $0^{+-}$			& $4.09\pm0.23$					& 	$10.17\pm0.22$						\\
 $2^{++}$			& $4.45\pm0.27$					& 	$10.64\pm0.33$						\\
 $1^{+-}$			& $4.53\pm0.23$					& 	$10.70\pm0.53$						\\
 $1^{++}$			& $5.06\pm0.44$					& 	$11.09\pm0.60$						\\
 $0^{++}$			& $5.34\pm0.45$					& 	$11.20\pm0.48$						\\
 $0^{--}$			& $5.51\pm0.50$					& 	$11.48\pm0.75$						\\
				&							& 									\\
  \hline
  \end{tabular}
\caption{Charmonium hybrid $\left(cg\bar{c}\right)$ and bottomonium hybrid $\left(bg\bar{b}\right)$ mass predictions for each $J^{PC}$ channel analyzed in Ref.~\cite{Chen:2013zia}.}
\label{masses}
\end{table}


\section{Conclusions}

A systematic QSR analysis of heavy quarkonium hybrids was conducted in Ref.~\cite{Chen:2013zia} and mass predictions for low lying heavy hybrid states were given. Inclusion of the dimension-six condensates was found to stabilize sum rules that were unstable in previous QSR studies of heavy hybrids~\cite{Govaerts:1984hc,Govaerts:1985fx,Govaerts:1986pp}. This allowed reliable mass predictions to be made. Updated mass predictions for charmonium hybrids and bottomonium hybrids are listed in Table~\ref{masses}. 

The QSR determined heavy hybrid masses form two supermultiplets: one containing the negative-parity $\left(0,1,2\right)^{-+},\,1^{--}$ states, and one containing the positive-parity $\left(0,1\right)^{+-},\,\left(0,1,2\right)^{++}$ states. In both the charmonium and bottomonium hybrid spectra, the states in the negative-parity supermultiplet are lighter than those in the positive-parity supermultiplet. Also, in both spectra, the $0^{--}$ state is much heavier than states in either supermultiplet. This supermultiplet structure is in agreement with results found using lattice QCD~\cite{Liu:2012ze} and the quasi-gluon model~\cite{Guo:2008yz}. However, the predicted masses for non-exotic $J^{PC}$ channels in Table~\ref{masses} are lower than those predicted in Refs.~\cite{Liu:2012ze,Guo:2008yz}. In Ref.~\cite{Chen:2013zia}, it was noted that the heavy hybrid currents~\eqref{currents} with non-exotic $J^{PC}$ can also couple to conventional heavy quarkonium states. A qualitative study of mixing between hybrids and quarkonia was performed in Ref.~\cite{Chen:2013zia}, showing that the non-exotic hybrid masses increase as mixing with quarkonia increases. Thus the non-exotic hybrid masses given in Table~\ref{masses} should be interpreted as lower bounds. However, heavy hybrids with exotic $J^{PC}$ cannot mix with convential heavy quarkonia. Therefore the $0^{--}$, $0^{+-}$, $1^{-+}$ and $2^{+-}$ exotic hybrid masses listed in Table~\ref{masses} are new QSR predictions for the masses of these states. Future work will include a full QSR analysis of mixing between non-exotic heavy hybrids and heavy quarkonia. 

The LHCb collaboration has shown that $X(3872)$ has $J^{PC}=1^{++}$~\cite{Aaij:2013zoa}. There have been many exotic interpretations of this state~\cite{Chen:2014fza,Bodwin:2013nua,Faccini:2012pj}, including the suggestion that it may be a charmonium hybrid~\cite{Li:2004sta}. However, the QSR mass prediction for the $1^{++}$ charmonium hybrid in Refs.~\cite{Harnett:2012gs,Chen:2013zia} is significantly higher than the mass of the $X(3872)$, disfavoring a pure charmonium hybrid interpretation of this state. 

The heavy quarkonium-like XYZ states present a tantalizing puzzle. It has been widely speculated that some of the XYZ states could be exotic hadrons, including heavy quarkonium hybrids. These could signal their presence indirectly through supernumerary states in non-exotic $J^{PC}$ channels, or, more directly, through the discovery of heavy quarkonium-like states with exotic $J^{PC}$. No such states have been observed to date. However, the quantum numbers of many XYZ states are still unknown. Therefore, it is possible that some of the XYZ states could be heavy quarkonium hybrids. The mass predictions given in Table~\ref{masses} and in Ref.~\cite{Chen:2013zia} will help in assessing this possibility.


\end{document}